\documentclass[pre,aps,showpacs,reprint,english]{revtex4-1}
\usepackage{amsmath,amssymb,amsfonts,amsthm,bm}
\usepackage{babel}

\begin{document}
\author{Xavier \surname{Lamy}}
\affiliation{Institut Camille Jordan,  CNRS UMR 5208, Universit\'e Lyon 1, 43 blvd. du 11 novembre 1918, F-69622 Villeurbanne cedex, France}

\title{A new light on the breaking of uniaxial symmetry in nematics}

\date{\today}

\begin{abstract}
Within the Landau-de Gennes theory of liquid crystals, we study theoretically the equilibrium configurations with uniaxial symmetry. For an arbitrary form of the bulk energy density, we show that energy minimizers among uniaxially symmetric configurations are usually not equilibrium configurations. It was known before that uniaxiality can \textbf{sometimes} be broken by energy minimizers. We prove here that this is \textbf{ always} the case in one or two dimensions. Even more we prove that equilibrium configurations (not only minimizers) \textbf{cannot be uniaxial} in one or two dimensions (unless they are constant). Namely, uniaxial equilibrium configurations with at least one translational invariance direction must have a uniform director field. In the case of a hybrid nematic cell, or of a capillary with radial anchoring, our result implies that biaxial escape always occurs, and that it does so not only in the core of a defect: uniaxial order is destroyed in the whole cell.
\end{abstract}


\maketitle

\section{Introduction}

Nematic liquid crystals are composed of rigid rod-like molecules which tend to align in a common preferred direction. For a macroscopic description of such orientational ordering, several continuum theories are available, relying on different order parameters.

The state of alignment can be simply characterized by a director field $\bm{n}$, corresponding to the local preferred direction of orientation. Within such a description, topological constraints may force the appearance of defects: regions where the director field is not continuous. 
To obtain a finer understanding of such regions, one needs to introduce a scalar order parameter $s$, corresponding to the degree of alignment along the director $\bm{n}$. 
However, the $(s,\bm{n})$ description only accounts for uniaxial nematics, which correspond to a symmetrical case of the more general biaxial nematic phase. To describe biaxial regions, a tensorial order parameter $\bm{Q}$ is needed. 
Biaxiality has been used to theoretically describe defect cores 
\cite{mkaddemgartland00, kraljvirgazumer99, penzenstadlertrebin89, sonnetkilianhess95,lucarey07a,lucarey07b} and material frustration \cite{palffymuhoraygartlandkelly94, bisigartlandrossovirga03,ambrozicbisivirga08}, and has been observed experimentally \cite{madsendingemansnakatasamulski04, acharyaprimakkumar04}.

The $(s,\bm{n})$ description can be viewed as a special case of the $\bm{Q}$-tensor description, restricted to $\bm{Q}$-tensors with uniaxial symmetry. In  physical systems presenting some symmetry, existence of symmetric equilibrium configurations is a common phenomenon: such configurations can be obtained by looking for a solution with a special symmetrical \textit{ansatz}. In some cases this phenomenon can be formalized mathematically as a `principle of symmetric criticality' \cite{palais79}.  
The purpose of this note is to investigate whether the same principle applies to uniaxial symmetry in nematic liquid crystals: do there exist uniaxial $Q$-tensor equilibrium configurations? or is the uniaxial symmetry spontaneously broken?

We consider a Landau-de Gennes free energy for the $\bm{Q}$-tensor under strong surface anchoring. We do not work with the usual four-terms expansion of the bulk energy density but with a general frame invariant bulk energy density. We are particularly interested in the case of one dimensional or two dimensional configurations: that is, configurations exhibiting translational invariance in at least one direction of space \cite{palffymuhoraygartlandkelly94, bisigartlandrossovirga03,ambrozicbisivirga08,kraljvirgazumer99, sonnetkilianhess95}. Our main result is the following: \textbf{even if the boundary conditions enhance uniaxial symmetry, the uniaxial order is destroyed in the whole system} (unless the director field is uniform). 

In particular, the present paper sheds a very new light on the phenomenon of `biaxial escape' \cite{sonnetkilianhess95}. Our result is \textbf{fundamentally different} from the previous related ones in the literature. In fact, biaxiality was always shown to occur by means of free energy comparison methods, while we \textbf{only rely on the equilibrium equations}. In particular our results hold for all metastable configurations. Moreover, the appearance of biaxiality was usually related to special values of parameters such as the temperature \cite{mkaddemgartland00} -- which affects the bulk equilibrium --, or the size of the system \cite{bisigartlandrossovirga03} -- which affects the director deformation. We show instead that biaxiality occurs for \textbf{any value of the temperature} (since the bulk energy density we work with is arbitrary) and \textbf{any kind of director deformation.}

The plan of the paper is the following. In Section~\ref{sm} we introduce the mathematical model describing orientational order. In Section~\ref{su} we derive the equilibrium equations for a configuration with uniaxial symmetry, and discuss the appearance of an extra equation corresponding to equilibrium with respect to symmetry-breaking perturbations. Section~\ref{s12} contains the main results of the paper, which we further discuss in the last section.

\section{Model} \label{sm}

The orientational order of a nematic state can be described by the order tensor $\bm{Q}$, a symmetric, traceless, second rank tensor \cite{degennes}. In the isotropic phase, $\bm{Q}=0$. The uniaxial phase corresponds to a $\bm{Q}$-tensor with two equal eigenvalues: $\bm{Q}$ can be written in the form
\begin{equation}\label{ansatz}
\bm{Q}=s\left(\bm{n}\otimes\bm{n}-\frac{1}{3}\bm{I}\right),
\end{equation}
where $s\in\mathbb{R}$ is the scalar order parameter, and the unit vector $\bm{n}$ is the director. The biaxial phase corresponds to all the eigenvalues of $\bm{Q}$ being different. Five degrees of freedom are needed to describe a biaxial state, instead of only three for a uniaxial state.

Under strong surface anchoring, the total free energy takes the form
\begin{equation*}
\mathcal{F}[Q]=\int (f_e + f_b) \; dV,
\end{equation*}
where $f_e$ is the elastic free energy density, and $f_b$ is the bulk energy density. For the elastic contribution we adopt the one-constant approximation
\begin{equation*}
f_e = \frac{L}{2}|\nabla Q|^2.
\end{equation*}
For the bulk contribution, however, we consider an arbitrary frame invariant function. Frame invariance implies that $f_b$ depends only on the invariants of $\bm{Q}$:
\begin{equation*}
f_b = \varphi\left(\mathrm{tr}(\bm{Q}^2),\mathrm{tr}(\bm{Q}^3)\right),
\end{equation*}
where $\varphi$ is a smooth function of two variables. The four-terms expansion for $f_b$, usually considered in the literature, corresponds to $\varphi(x,y)=\frac{a}{2} x - \frac{b}{3} y + \frac{c}{4}x^2$.

At equilibrium the first variation of the free energy should vanish:
\begin{equation*}
\delta\mathcal{F}[\bm{Q}]=0.
\end{equation*}
These equilibrium equations for $\mathcal{F}$ are straightforward to compute, and read
\begin{equation}\label{LG}
L\Delta \bm{Q} = 2\partial_1\varphi\: \bm{Q} + 3 \partial_2\varphi\: \left(\bm{Q}^2-\frac{|\bm{Q}|^2}{3}\bm{I}\right).
\end{equation}
Here, $\bm{I}$ denotes the identity tensor, and $|\bm{Q}|^2=\mathrm{tr}(\bm{Q}^2)$.

We are going to investigate the existence of solutions of the equilibrium equations \eqref{LG} which satisfy in addition the uniaxial symmetry constraint \eqref{ansatz}. In the next section we derive the equations for such a configuration, showing that a configuration which is an equilibrium under the constraint of uniaxial symmetry, need not, in general, be an unconstrained equilibrium. In fact an additional equation needs to be satisfied by the director field.

\section{Uniaxial equilibrium} \label{su}

We want to understand under which conditions a solution to the equilibrium equations \eqref{LG} can be written in the form
\eqref{ansatz}
for some scalar field $s$ and unit vector field $\bm{n}$.

Let us remark here that the spherically symmetric radial hedgehog \cite{mkaddemgartland00} provides an example of uniaxial equilibrium. However, in the particular case of the radial hedgehog, uniaxial symmetry is a consequence of spherical symmetry, for which the principle of symmetric criticality applies \cite{palais79}. In general we can not use such an argument.

Henceforth we assume that $\bm{Q}$ is a uniaxial equilibrium: it satisfies \eqref{LG} and \eqref{ansatz}.
Plugging the \textit{ansatz} \eqref{ansatz} into \eqref{LG}, we obtain, after rearranging the terms,
\begin{equation}\label{LGU}
\begin{split}
&\left[ \Delta s - 3|\nabla \bm{n}|^2 s -\frac{1}{L}(2s\partial_1\varphi + s^2 \partial_2\varphi)\right]\left(\bm{n}\otimes \bm{n}-\frac{1}{3} I\right) \\
& + 2\bm{n} \odot (s\Delta \bm{n} + 2(\nabla s\cdot \nabla)\bm{n} + s|\nabla \bm{n}|^2 \bm{n}) \\
& + 2 s \left[ \sum_k \partial_k \bm{n} \otimes\partial_k \bm{n} + |\nabla \bm{n}|^2 \left(  \bm{n}\otimes \bm{n}-\bm{I}\right) \right] =0.
\end{split}
\end{equation}
Here $\odot$ denotes the symmetric tensor product: the $(i,j)$ component of $\bm{n}\odot\bm{m}$ is $(\bm{n}_i\bm{m}_j+\bm{n}_j\bm{m}_i)/2$.

Differentiation of the constraint $|\bm{n}|^2=1$ implies $\bm{n}\cdot\partial_k\bm{n}=0$. Using this orthogonality property, it is straightforward to check that the three lines in \eqref{LGU} are orthogonal to each other (for the usual scalar product on square matrices). Therefore they must all vanish, which implies:
\begin{equation}\label{U}
\left\lbrace
\begin{gathered}
\Delta s = 3 |\nabla \bm{n}|^2 s +\frac{1}{L}(2s\partial_1\varphi + s^2 \partial_2\varphi), \\
s\Delta \bm{n} + 2 (\nabla s \cdot\nabla)\bm{n} = -s|\nabla \bm{n}|^2 \bm{n},
\end{gathered}
\right.
\end{equation}
and 
\begin{equation}\label{extra}
2 \sum_{k=1}^3 \partial_k \bm{n} \otimes \partial_k \bm{n}  = |\nabla \bm n|^2 \left(\bm{I}-\bm{n} \otimes \bm{n}\right),
\end{equation}
in regions where $\bm{Q}\neq 0$.

Let us remark here that \eqref{U} corresponds to $(s,\bm{n})$ being an equilibrium for the free energy
\begin{equation*}
\begin{split}
& F[s,\bm{n}]  =\mathcal{F}[\bm{Q}]\\
& =\int \left[\frac{L}{2}\left(\frac{2}{3}|\nabla s|^2 + 2s^2|\nabla \bm{n}|^2\right)+\varphi(2s^2/3,2s^3/9) \right] dx,
\end{split}\end{equation*}
under the constraint $|\bm{n}|=1$. Hence \eqref{U} simply tells us that $\bm{Q}$ is an equilibrium for $\mathcal{F}$ \textit{under the constraint of uniaxial symmetry}. Of course, this is indeed the case, since we have assumed that $\bm{Q}$ is an equilibrium without constraint. 

In addition to the equations \eqref{U}, we obtain the extra equation \eqref{extra}.
Therefore, being an unconstrained equilibrium is really stronger than being an equilibrium under the uniaxial symmetry constraint: if a configuration is an equilibrium with respect to perturbations preserving the uniaxial symmetry, it may not be an equilibrium with respect to arbitrary perturbations. Hence the principle of symmetric criticality is broken here. The additional equation \eqref{extra} can be interpreted as a necessary condition for the vanishing of the free energy's first variation when the perturbation breaks the uniaxial symmetry.

\section{In one and two dimensions}\label{s12}

Here we derive consequences of \eqref{U} and \eqref{extra} for configurations depending on only one or two directions in space. 
Such a symmetry assumption is actually relevant for many nematic systems that are interesting both theoretically and for application purposes. For instance, in nematic cells bounded by two parallel plates with competing anchoring, one usually looks for one dimensional solutions \cite{palffymuhoraygartlandkelly94, bisigartlandrossovirga03,ambrozicbisivirga08}. Such hybrid nematic cells provide a model system for understanding the physics of frustration, and this kind of geometry occurs in several nematic based optical devices. 
Another relevant geometry is the cylindrical one, in which two dimensional configurations can be considered \cite{ kraljvirgazumer99, sonnetkilianhess95,lucarey07a,lucarey07b}, with applications to high performance fibers \cite{cheongrey04,chanetal05,jianhurtsheldoncrawford06}.

Our conclusion is that such configurations must have a uniform director field. Of course the one dimensional case is contained in the two dimensional case as a particular case. We nevertheless treat it separately, since the argument is much simpler.

\subsection{The one dimensional case}

Here we consider a uniaxial equilibrium configuration depending only on one direction in space. We may assume $\partial_2 \bm{n} =\partial_3 \bm{n}=0$, since the free energy is frame invariant. We deduce, as a direct consequence of \eqref{extra}, that $\partial_1 \bm{n}=0$. As a matter of fact, applying \eqref{extra} -- which is an equality of matrices -- to the vector $\partial_1\bm{n}$ yields
\begin{equation*}
|\partial_1\bm{n}|^2 \partial_1\bm{n}=0.
\end{equation*}
Therefore the director field must be uniform.

\subsection{The two dimensional case}

For two dimensional configurations, \eqref{extra} does not directly force $n$ to be uniform. In fact, a cylindrically symmetric director field studied by Cladis and Kl\'e{}man \cite{cladiskleman72}, which is of the form
\begin{equation*}
\bm{n}=\cos \varphi(r) \:\bm{e_r} + \sin \varphi(r) \:\bm{e_z}\quad\text{with }r\frac{d\varphi}{dr} = \cos\varphi,
\end{equation*}
satisfies \eqref{extra}. However, there can not exist any scalar field $s$ satisfying \eqref{U} for this particular choice of $\bm{n}$.

In general, we will show that for a two dimensional configurations, \eqref{extra} imposes very strong conditions on the director field $\bm{n}$. These conditions turn out to be incompatible with the existence of a scalar field $s$ satisfying \eqref{U}.

Let us consider a two dimensional equilibrium. Using frame invariance, we may assume $\partial_3 \bm{n}=0$. Applying the matrix equality \eqref{extra} to the vectors $\partial_1\bm{n}$, respectively $\partial_2\bm{n}$, we find two vector equalities. 
After taking the scalar product of these two equalities with both $\partial_1\bm{n}$ and $\partial_2\bm{n}$, we are finally led to
\begin{equation}\label{conseq2d}
\partial_1 \bm{n} \cdot \partial_2\bm{n} = 0\quad\text{and}\quad |\partial_1 \bm{n}|^2=|\partial_2 \bm{n}|^2.
\end{equation}
These conditions turn out to be incompatible with the existence of a scalar field $s$ satisfying \eqref{U}, unless the director field $\bm{n}$ is uniform. 

In fact, a straightforward consequence  obtained by differentiating the equations \eqref{conseq2d}, is that $\Delta \bm{n}$ is orthogonal to $\partial_1\bm{n}$ and $\partial_2\bm{n}$. On the other hand, recall that the constraint $|\bm{n}|=1$ forces $\bm{n}$ to be orthogonal to $\partial_1 \bm{n}$ and $\partial_2 \bm{n}$. Hence, taking the scalar product of the second equation in \eqref{U} with $\partial_1\bm{n}$ and $\partial_2\bm{n}$, we are left with
\begin{equation*}
\partial_1 s |\nabla \bm{n}|^2 = \partial_2 s |\nabla \bm{n}|^2 = 0.
\end{equation*}
In other words, the scalar field $s$ must be constant in every region where the director field $\bm{n}$ is not uniform.

To conclude, we just need to remark that, in a region where the scalar field $s$ is constant, and non zero, the first equation in \eqref{U} implies that $\nabla \bm{n}$ has constant norm. The geometry of the unit sphere, where $\bm{n}$ takes its values, makes it impossible: we may work with rescaled variables and assume $|\nabla\bm{n}|^2=2$, so that $\bm{n}$ satisfies
\begin{equation*}
\partial_1 \bm{n}\cdot\partial_2\bm{n}=0 \quad\text{and}\quad |\partial_1 \bm{n}|^2=|\partial_2 \bm{n}|^2=1.
\end{equation*}
Therefore $\bm{n}$ induces locally an isometry between the Euclidean plane -- which has zero curvature -- and the unit sphere -- which has positive curvature. By Gauss' \textit{Theorema egregium}, no such isometry can exist. Hence, a region where $\bm{n}$ is not uniform can not exist: the director field $\bm{n}$ is uniform everywhere. 

In fact we obtain even a stronger result: a general two dimensional equilibrium configuration can not be uniaxial, in any -- even small -- open region, unless it has constant director. Indeed, the map $\bm{Q}$ is analytical (i.e. it is locally the sum of a converging power series) \cite[Theorem 6.7.6]{morrey}. As a consequence, the region where it has uniaxial symmetry must be either the whole domain, or be negligeable \cite[Proposition 14]{majumdarzarnescu10}. Our argument shows that it is not the whole domain, hence it is negligeable.

\section{Conclusions}

We have shown that, for a nematic equilibrium configuration presenting translational invariance in one direction, there are only two options: either it does not have any regions with uniaxial symmetry, or it has uniform director field. In particular, when the boundary conditions prevent the director field from being uniform, as it is the case in hybrid cells or in capillaries with radial anchoring, then at equilibrium uniaxial order is destroyed spontaneously within the whole system. In other words, for translationally invariant configurations, biaxial escape has to occur.

Biaxiality had in fact been predicted in such geometries \cite{sonnetkilianhess95, palffymuhoraygartlandkelly94, bisigartlandrossovirga03}, but it was supposed to stay confined to small regions, and to occur only in some parameter range. Here we have provided a rigorous proof that biaxiality must occur everywhere, and for any values of the parameter: the configurations interpreted as uniaxial just correspond to a small degree of biaxiality. Our proof does not rely on free energy minimization, but only on the equilibrium equations -- in particular it affects all metastable configurations.  It is also remarkable that our results do not depend on the form of the bulk energy density, whereas all the previously cited workers used a four-terms approximation.

After this work was completed, G. Napoli brought the article \cite{biscarinapoliturzi06} to our attention. In \cite{biscarinapoliturzi06} results similar to the ones in Section~\ref{s12} of the present paper are stated under the additional assumption that the configuration is energy minimizing. However, we were not able to follow completely the arguments in \cite{biscarinapoliturzi06}.

\begin{acknowledgments}
We thank Petru Mironescu for stimulating discussions and Maxime Ignacio for his useful comments.
\end{acknowledgments}

\bibliographystyle{apsrev4-1}
\bibliography{uniax}

\begin{thebibliography}{10}%
\makeatletter
\providecommand \@ifxundefined [1]{%
 \ifx #1\undefined \expandafter \@firstoftwo
 \else \expandafter \@secondoftwo
\fi
}%
\providecommand \@ifnum [1]{%
 \ifnum #1\expandafter \@firstoftwo
 \else \expandafter \@secondoftwo
\fi
}%
\providecommand \enquote [1]{``#1''}%
\providecommand \bibnamefont  [1]{#1}%
\providecommand \bibfnamefont [1]{#1}%
\providecommand \citenamefont [1]{#1}%
\providecommand\href[0]{\@sanitize\@href}%
\providecommand\@href[1]{\endgroup\@@startlink{#1}\endgroup\@@href}%
\providecommand\@@href[1]{#1\@@endlink}%
\providecommand \@sanitize [0]{\begingroup\catcode`\&12\catcode`\#12\relax}%
\@ifxundefined \pdfoutput {\@firstoftwo}{%
 \@ifnum{\z@=\pdfoutput}{\@firstoftwo}{\@secondoftwo}%
}{%
 \providecommand\@@startlink[1]{\leavevmode\special{html:<a href="#1">}}%
 \providecommand\@@endlink[0]{\special{html:</a>}}%
}{%
 \providecommand\@@startlink[1]{%
  \leavevmode
  \pdfstartlink
   attr{/Border[0 0 1 ]/H/I/C[0 1 1]}%
   user{/Subtype/Link/A<</Type/Action/S/URI/URI(#1)>>}%
  \relax
 }%
 \providecommand\@@endlink[0]{\pdfendlink}%
}%
\providecommand \url  [0]{\begingroup\@sanitize \@url }%
\providecommand \@url [1]{\endgroup\@href {#1}{\urlprefix}}%
\providecommand \urlprefix [0]{URL }%
\providecommand \Eprint[0]{\href }%
\@ifxundefined \urlstyle {%
  \providecommand \doi [1]{doi:\discretionary{}{}{}#1}%
}{%
  \providecommand \doi [0]{doi:\discretionary{}{}{}\begingroup
  \urlstyle{rm}\Url }%
}%
\providecommand \doibase [0]{http://dx.doi.org/}%
\providecommand \Doi[1]{\href{\doibase#1}}%
\providecommand \bibAnnote [3]{%
  \BibitemShut{#1}%
  \begin{quotation}\noindent
    \textsc{Key:}\ #2\\\textsc{Annotation:}\ #3%
  \end{quotation}%
}%
\providecommand \bibAnnoteFile [2]{%
  \IfFileExists{#2}{\bibAnnote {#1} {#2} {\input{#2}}}{}%
}%
\providecommand \typeout [0]{\immediate \write \m@ne }%
\providecommand \selectlanguage [0]{\@gobble}%
\providecommand \bibinfo [0]{\@secondoftwo}%
\providecommand \bibfield [0]{\@secondoftwo}%
\providecommand \translation [1]{[#1]}%
\providecommand \BibitemOpen[0]{}%
\providecommand \bibitemStop [0]{}%
\providecommand \bibitemNoStop [0]{.\EOS\space}%
\providecommand \EOS [0]{\spacefactor3000\relax}%
\providecommand \BibitemShut [1]{\csname bibitem#1\endcsname}%
\bibitem{mkaddemgartland00}%
  \BibitemOpen
  \bibfield{author}{%
  \bibinfo {author} {\bibfnamefont{S.}~\bibnamefont{Mkaddem}}\ and\ \bibinfo
  {author} {\bibfnamefont{E.~C.}\ \bibnamefont{Gartland}},\ }%
  \bibfield{journal}{%
  \Doi{10.1103/PhysRevE.62.6694}{\bibinfo {journal} {Phys. Rev. E}}\ }%
  \textbf{\bibinfo {volume} {62}},\ \bibinfo {pages} {6694} (\bibinfo {month}
  {Nov}\ \bibinfo {year} {2000}),\
  \url{http://link.aps.org/doi/10.1103/PhysRevE.62.6694}%
  \bibAnnoteFile{NoStop}{mkaddemgartland00}%
\bibitem{kraljvirgazumer99}%
  \BibitemOpen
  \bibfield{author}{%
  \bibinfo {author} {\bibfnamefont{S.}~\bibnamefont{Kralj}}, \bibinfo {author}
  {\bibfnamefont{E.~G.}\ \bibnamefont{Virga}},\ and\ \bibinfo {author}
  {\bibfnamefont{S.}~\bibnamefont{\ifmmode~\check{Z}\else \v{Z}\fi{}umer}},\ }%
  \bibfield{journal}{%
  \Doi{10.1103/PhysRevE.60.1858}{\bibinfo {journal} {Phys. Rev. E}}\ }%
  \textbf{\bibinfo {volume} {60}},\ \bibinfo {pages} {1858} (\bibinfo {month}
  {Aug}\ \bibinfo {year} {1999}),\
  \url{http://link.aps.org/doi/10.1103/PhysRevE.60.1858}%
  \bibAnnoteFile{NoStop}{kraljvirgazumer99}%
\bibitem{penzenstadlertrebin89}%
  \BibitemOpen
  \bibfield{author}{%
  \bibinfo {author} {\bibfnamefont{E.}~\bibnamefont{Penzenstadler}}\ and\
  \bibinfo {author} {\bibfnamefont{H.-R.}\ \bibnamefont{Trebin}},\ }%
  \bibfield{journal}{%
  \Doi{10.1051/jphys:019890050090102700}{\bibinfo {journal} {J. Phys. France}}\
  }%
  \textbf{\bibinfo {volume} {50}},\ \bibinfo {pages} {1027} (\bibinfo {year}
  {1989}),\ \url{http://dx.doi.org/10.1051/jphys:019890050090102700}%
  \bibAnnoteFile{NoStop}{penzenstadlertrebin89}%
\bibitem{sonnetkilianhess95}%
  \BibitemOpen
  \bibfield{author}{%
  \bibinfo {author} {\bibfnamefont{A.}~\bibnamefont{Sonnet}}, \bibinfo {author}
  {\bibfnamefont{A.}~\bibnamefont{Kilian}},\ and\ \bibinfo {author}
  {\bibfnamefont{S.}~\bibnamefont{Hess}},\ }%
  \bibfield{journal}{%
  \Doi{10.1103/PhysRevE.52.718}{\bibinfo {journal} {Phys. Rev. E}}\ }%
  \textbf{\bibinfo {volume} {52}},\ \bibinfo {pages} {718} (\bibinfo {month}
  {Jul}\ \bibinfo {year} {1995}),\
  \url{http://link.aps.org/doi/10.1103/PhysRevE.52.718}%
  \bibAnnoteFile{NoStop}{sonnetkilianhess95}%
\bibitem{lucarey07a}%
  \BibitemOpen
  \bibfield{author}{%
  \bibinfo {author} {\bibfnamefont{G.~D.}\ \bibnamefont{Luca}}\ and\ \bibinfo
  {author} {\bibfnamefont{A.~D.}\ \bibnamefont{Rey}},\ }%
  \bibfield{journal}{%
  \Doi{10.1063/1.2711436}{\bibinfo {journal} {J. Chem. Phys.}}\ }%
  \textbf{\bibinfo {volume} {126}},\ \bibinfo {eid} {094907} (\bibinfo {year}
  {2007}),\ \url{http://link.aip.org/link/?JCP/126/094907/1}%
  \bibAnnoteFile{NoStop}{lucarey07a}%
\bibitem{lucarey07b}%
  \BibitemOpen
  \bibfield{author}{%
  \bibinfo {author} {\bibfnamefont{G.~D.}\ \bibnamefont{Luca}}\ and\ \bibinfo
  {author} {\bibfnamefont{A.~D.}\ \bibnamefont{Rey}},\ }%
  \bibfield{journal}{%
  \Doi{10.1063/1.2775451}{\bibinfo {journal} {J. Chem. Phys.}}\ }%
  \textbf{\bibinfo {volume} {127}},\ \bibinfo {eid} {104902} (\bibinfo {year}
  {2007}),\ \url{http://link.aip.org/link/?JCP/127/104902/1}%
  \bibAnnoteFile{NoStop}{lucarey07b}%
\bibitem{palffymuhoraygartlandkelly94}%
  \BibitemOpen
  \bibfield{author}{%
  \bibinfo {author} {\bibfnamefont{P.}~\bibnamefont{Palffy-Muhoray}}, \bibinfo
  {author} {\bibfnamefont{E.}~\bibnamefont{Gartland}},\ and\ \bibinfo {author}
  {\bibfnamefont{J.}~\bibnamefont{Kelly}},\ }%
  \bibfield{journal}{%
  \Doi{10.1080/02678299408036543}{\bibinfo {journal} {Liquid Crystals}}\ }%
  \textbf{\bibinfo {volume} {16}},\ \bibinfo {pages} {713} (\bibinfo {year}
  {1994}),\ \url{http://www.tandfonline.com/doi/abs/10.1080/02678299408036543}%
  \bibAnnoteFile{NoStop}{palffymuhoraygartlandkelly94}%
\bibitem{bisigartlandrossovirga03}%
  \BibitemOpen
  \bibfield{author}{%
  \bibinfo {author} {\bibfnamefont{F.}~\bibnamefont{Bisi}}, \bibinfo {author}
  {\bibfnamefont{E.~C.}\ \bibnamefont{Gartland}}, \bibinfo {author}
  {\bibfnamefont{R.}~\bibnamefont{Rosso}},\ and\ \bibinfo {author}
  {\bibfnamefont{E.~G.}\ \bibnamefont{Virga}},\ }%
  \bibfield{journal}{%
  \Doi{10.1103/PhysRevE.68.021707}{\bibinfo {journal} {Phys. Rev. E}}\ }%
  \textbf{\bibinfo {volume} {68}},\ \bibinfo {pages} {021707} (\bibinfo {month}
  {Aug}\ \bibinfo {year} {2003}),\
  \url{http://link.aps.org/doi/10.1103/PhysRevE.68.021707}%
  \bibAnnoteFile{NoStop}{bisigartlandrossovirga03}%
\bibitem{ambrozicbisivirga08}%
  \BibitemOpen
  \bibfield{author}{%
  \bibinfo {author} {\bibfnamefont{M.}~\bibnamefont{Ambro{\v{z}}i{\v{c}}}},
  \bibinfo {author} {\bibfnamefont{F.}~\bibnamefont{Bisi}},\ and\ \bibinfo
  {author} {\bibfnamefont{E.~G.}\ \bibnamefont{Virga}},\ }%
  \bibfield{journal}{%
  \Doi{10.1007/s00161-008-0077-x}{\bibinfo {journal} {Continuum Mech.
  Thermodyn.}}\ }%
  \textbf{\bibinfo {volume} {20}},\ \bibinfo {pages} {193} (\bibinfo {year}
  {2008}),\ \url{http://dx.doi.org/10.1007/s00161-008-0077-x}%
  \bibAnnoteFile{NoStop}{ambrozicbisivirga08}%
\bibitem{madsendingemansnakatasamulski04}%
  \BibitemOpen
  \bibfield{author}{%
  \bibinfo {author} {\bibfnamefont{L.~A.}\ \bibnamefont{Madsen}}, \bibinfo
  {author} {\bibfnamefont{T.~J.}\ \bibnamefont{Dingemans}}, \bibinfo {author}
  {\bibfnamefont{M.}~\bibnamefont{Nakata}},\ and\ \bibinfo {author}
  {\bibfnamefont{E.~T.}\ \bibnamefont{Samulski}},\ }%
  \bibfield{journal}{%
  \Doi{10.1103/PhysRevLett.92.145505}{\bibinfo {journal} {Phys. Rev. Lett.}}\
  }%
  \textbf{\bibinfo {volume} {92}},\ \bibinfo {pages} {145505} (\bibinfo {month}
  {Apr}\ \bibinfo {year} {2004}),\
  \url{http://link.aps.org/doi/10.1103/PhysRevLett.92.145505}%
  \bibAnnoteFile{NoStop}{madsendingemansnakatasamulski04}%
\bibitem{acharyaprimakkumar04}%
  \BibitemOpen
  \bibfield{author}{%
  \bibinfo {author} {\bibfnamefont{B.~R.}\ \bibnamefont{Acharya}}, \bibinfo
  {author} {\bibfnamefont{A.}~\bibnamefont{Primak}},\ and\ \bibinfo {author}
  {\bibfnamefont{S.}~\bibnamefont{Kumar}},\ }%
  \bibfield{journal}{%
  \Doi{10.1103/PhysRevLett.92.145506}{\bibinfo {journal} {Phys. Rev. Lett.}}\
  }%
  \textbf{\bibinfo {volume} {92}},\ \bibinfo {pages} {145506} (\bibinfo {month}
  {Apr}\ \bibinfo {year} {2004}),\
  \url{http://link.aps.org/doi/10.1103/PhysRevLett.92.145506}%
  \bibAnnoteFile{NoStop}{acharyaprimakkumar04}%
\bibitem{palais79}%
  \BibitemOpen
  \bibfield{author}{%
  \bibinfo {author} {\bibfnamefont{R.}~\bibnamefont{Palais}},\ }%
  \bibfield{journal}{%
  \Doi{10.1007/BF01941322}{\bibinfo {journal} {Comm. Math. Phys.}}\ }%
  \textbf{\bibinfo {volume} {69}},\ \bibinfo {pages} {19} (\bibinfo {year}
  {1979}),\ ISSN \bibinfo {issn} {0010-3616},\
  \url{http://dx.doi.org/10.1007/BF01941322}%
  \bibAnnoteFile{NoStop}{palais79}%
\bibitem{degennes}%
  \BibitemOpen
  \bibfield{author}{%
  \bibinfo {author} {\bibfnamefont{P.}~\bibnamefont{De~Gennes}}\ and\ \bibinfo
  {author} {\bibfnamefont{J.}~\bibnamefont{Prost}},\ }%
  \emph{\bibinfo {title} {The {P}hysics of {L}iquid {C}rystals}},\ \bibinfo
  {edition} {2nd}\ ed.\ (\bibinfo {publisher} {Oxford {U}niversity {P}ress},\
  \bibinfo {year} {1993})%
  \bibAnnoteFile{NoStop}{degennes}%
\bibitem{cheongrey04}%
  \BibitemOpen
  \bibfield{author}{%
  \bibinfo {author} {\bibfnamefont{A.-G.}\ \bibnamefont{Cheong}}\ and\ \bibinfo
  {author} {\bibfnamefont{A.~D.}\ \bibnamefont{Rey}},\ }%
  \bibfield{journal}{%
  \Doi{10.1080/02678290412331282109}{\bibinfo {journal} {Liquid crystals}}\ }%
  \textbf{\bibinfo {volume} {31}},\ \bibinfo {pages} {1271} (\bibinfo {year}
  {2004}),\
  \url{http://www.tandfonline.com/doi/abs/10.1080/02678290412331282109}%
  \bibAnnoteFile{NoStop}{cheongrey04}%
\bibitem{chanetal05}%
  \BibitemOpen
  \bibfield{author}{%
  \bibinfo {author} {\bibfnamefont{C.}~\bibnamefont{Chan}}, \bibinfo {author}
  {\bibfnamefont{G.}~\bibnamefont{Crawford}}, \bibinfo {author}
  {\bibfnamefont{Y.}~\bibnamefont{Gao}}, \bibinfo {author}
  {\bibfnamefont{R.}~\bibnamefont{Hurt}}, \bibinfo {author}
  {\bibfnamefont{K.}~\bibnamefont{Jian}}, \bibinfo {author}
  {\bibfnamefont{H.}~\bibnamefont{Li}}, \bibinfo {author}
  {\bibfnamefont{B.}~\bibnamefont{Sheldon}}, \bibinfo {author}
  {\bibfnamefont{M.}~\bibnamefont{Sousa}},\ and\ \bibinfo {author}
  {\bibfnamefont{N.}~\bibnamefont{Yang}},\ }%
  \bibfield{journal}{%
  \Doi{http://dx.doi.org/10.1016/j.carbon.2005.04.033}{\bibinfo {journal}
  {Carbon}}\ }%
  \textbf{\bibinfo {volume} {43}},\ \bibinfo {pages} {2431 } (\bibinfo {year}
  {2005}),\ ISSN \bibinfo {issn} {0008-6223},\
  \url{http://www.sciencedirect.com/science/article/pii/S0008622305002514}%
  \bibAnnoteFile{NoStop}{chanetal05}%
\bibitem{jianhurtsheldoncrawford06}%
  \BibitemOpen
  \bibfield{author}{%
  \bibinfo {author} {\bibfnamefont{K.}~\bibnamefont{Jian}}, \bibinfo {author}
  {\bibfnamefont{R.~H.}\ \bibnamefont{Hurt}}, \bibinfo {author}
  {\bibfnamefont{B.~W.}\ \bibnamefont{Sheldon}},\ and\ \bibinfo {author}
  {\bibfnamefont{G.~P.}\ \bibnamefont{Crawford}},\ }%
  \bibfield{journal}{%
  \Doi{10.1063/1.2197319}{\bibinfo {journal} {Applied Physics Letters}}\ }%
  \textbf{\bibinfo {volume} {88}},\ \bibinfo {eid} {163110} (\bibinfo {year}
  {2006}),\ \url{http://link.aip.org/link/?APL/88/163110/1}%
  \bibAnnoteFile{NoStop}{jianhurtsheldoncrawford06}%
\bibitem{cladiskleman72}%
  \BibitemOpen
  \bibfield{author}{%
  \bibinfo {author} {\bibfnamefont{P.}~\bibnamefont{Cladis}}\ and\ \bibinfo
  {author} {\bibfnamefont{M.}~\bibnamefont{Kl\'eman}},\ }%
  \bibfield{journal}{%
  \bibinfo {journal} {J. Phys. France}\ }%
  \textbf{\bibinfo {volume} {33}} (\bibinfo {year} {1972})%
  \bibAnnoteFile{NoStop}{cladiskleman72}%
\bibitem{morrey}%
  \BibitemOpen
  \bibfield{author}{%
  \bibinfo {author} {\bibfnamefont{C.}~\bibnamefont{Morrey~Jr}},\ }%
  \emph{\bibinfo {title} {Multiple integrals in the calculus of variations}}\
  (\bibinfo {publisher} {Springer},\ \bibinfo {year} {1966})%
  \bibAnnoteFile{NoStop}{morrey}%
\bibitem{majumdarzarnescu10}%
  \BibitemOpen
  \bibfield{author}{%
  \bibinfo {author} {\bibfnamefont{A.}~\bibnamefont{Majumdar}}\ and\ \bibinfo
  {author} {\bibfnamefont{A.}~\bibnamefont{Zarnescu}},\ }%
  \bibfield{journal}{%
  \Doi{10.1007/s00205-009-0249-2}{\bibinfo {journal} {Arch. Ration. Mech.
  Anal.}}\ }%
  \textbf{\bibinfo {volume} {196}},\ \bibinfo {pages} {227} (\bibinfo {year}
  {2010}),\ \url{http://dx.doi.org/10.1007/s00205-009-0249-2}%
  \bibAnnoteFile{NoStop}{majumdarzarnescu10}%
\bibitem{biscarinapoliturzi06}%
  \BibitemOpen
  \bibfield{author}{%
  \bibinfo {author} {\bibfnamefont{P.}~\bibnamefont{Biscari}}, \bibinfo
  {author} {\bibfnamefont{G.}~\bibnamefont{Napoli}},\ and\ \bibinfo {author}
  {\bibfnamefont{S.}~\bibnamefont{Turzi}},\ }%
  \bibfield{journal}{%
  \Doi{10.1103/PhysRevE.74.031708}{\bibinfo {journal} {Phys. Rev. E}}\ }%
  \textbf{\bibinfo {volume} {74}},\ \bibinfo {pages} {031708} (\bibinfo {month}
  {Sep}\ \bibinfo {year} {2006}),\
  \url{http://link.aps.org/doi/10.1103/PhysRevE.74.031708}%
  \bibAnnoteFile{NoStop}{biscarinapoliturzi06}%
\end{thebibliography}%

\end{document}